\documentclass[12pt]{article}
\usepackage{times}
\usepackage{geometry}
\usepackage[utf8]{inputenc}
\usepackage{enumitem,amssymb}
\usepackage{ragged2e}
\newlist{thematic}{itemize}{8}
\setlist[thematic]{label=$\square$}
\usepackage{pifont}

\def\lsim{\lower 2pt \hbox{$\, \buildrel {\scriptstyle <}\over
         {\scriptstyle \sim}\,$}}
\newcommand\gsim{\buildrel > \over \sim}

\usepackage{wrapfig}
\usepackage{graphicx}

\begin{document}
\raggedright
\huge
Astro2020 Science White Paper \linebreak

Prospects for Pulsar Studies at MeV Energies \linebreak
\normalsize

\noindent \textbf{Thematic Areas:} \hspace*{60pt} $\square$ Planetary Systems \hspace*{10pt} $\square$ Star and Planet Formation \hspace*{20pt}\linebreak
$\square\llap {X}$ Formation and Evolution of Compact Objects \hspace*{31pt} $\square$ Cosmology and Fundamental Physics \linebreak
  $\square$  Stars and Stellar Evolution \hspace*{1pt} $\square$ Resolved Stellar Populations and their Environments \hspace*{40pt} \linebreak
  $\square$    Galaxy Evolution   \hspace*{45pt} $\square$             Multi-Messenger Astronomy and Astrophysics \hspace*{65pt} \linebreak
  
\textbf{Principal Author:}

Name:	Alice K. Harding
 \linebreak						
Institution:  NASA Goddard Space Flight Center
 \linebreak
Email: Alice.K.Harding@nasa.gov
 \linebreak
Phone:  (301) 286-7824
 \linebreak
 
\textbf{Co-authors:} 
Matthew Kerr, Naval Research Laboratory, Washington, DC; matthew.kerr@nrl.navy.mil

Marco Ajello, Clemson University

Denis Bernard, LLR, Ecole Polytechnique, CNRS/IN2P3, France

Harsha Blumer, West Virginia University

Isabelle Grenier, Université Paris Diderot/CEA Saclay, France

Sylvain Guiriec, George Washington University/NASA Goddard Space Flight Center

Francesco Longo, University and INFN, Trieste

Antonios Manousakis, University of Sharjah, Sharjah, UAE

Pablo Saz-Parkinson, University of Hong Kong, Santa Cruz Institute for Particle Physics

Chanda Prescod-Weinstein, University of New Hampshire

Zorawar Wadiasingh, NASA Goddard Space Flight Center, USRA/NPP Fellow

George Younes, George Washington University

Silvia Zane, University College London

Bing Zhang, University of Nevada-Las Vegas
\linebreak

\textbf{Abstract}
Enabled by the \textit{Fermi} Large Area Telescope, we now know young and recycled pulsars fill the gamma-ray sky, and we are beginning to understand their emission mechanism and their distribution throughout the Galaxy. However, key questions remain: Is there a large population of pulsars near the Galactic center? Why do the most energetic pulsars shine so brightly in MeV gamma rays but not always at GeV energies? What is the source and nature of the pair plasma in pulsar magnetospheres, and what role does the polar cap accelerator play? Addressing these questions calls for a sensitive, wide-field MeV telescope, which can detect the population of MeV-peaked pulsars hinted at by Fermi and hard X-ray telescopes and characterize their spectral shape and polarization.

\pagebreak
\section{Introduction}

Our understanding of rotation-powered pulsars (RPP) has undergone a revolution over the past ten years with the \textit{Fermi} Gamma-Ray Space Telescope enabling discoveries of more than 230 new $\gamma$-ray pulsars \cite{Abdo2013,pulsar_list} above 100 MeV.  Although ten times as many RPPs have been discovered in the radio, the high-energy emission from $\gamma$-ray pulsars more directly reveals the particle production and acceleration in their magnetospheres and dominates the energy budget.\\

Recent rapid advances in global models of the pulsar magnetosphere \cite{Cerutti2016,Kala2018} suggest that the emission in the {\it Fermi} band comes from the most energetic particles that are accelerated in the current sheet outside the light cylinder.  But, to account for the narrow peaks of the observed light curves, these models require a source of abundant electron-positron pairs to screen the large electric fields and establish a near force-free magnetosphere.  Many emission models suggest that these pairs produce a synchrotron radiation component whose spectral power peaks in the 0.1--10 MeV range.  Detection of this radiation can give us crucial information on the origin and spectrum of the pairs.  Unfortunately, the emission in this band has been detected for only a few of the youngest RPPs, notably the Crab and PSR B1509-58.  Hard X-ray observations have detected a population of young RPPs exhibiting hard, non-thermal spectra extending from 20 to 200 keV, strongly implying that their spectral power may peak at MeV energies.  Most of these, a group of 10 ``MeV pulsars", have pulsed detection in neither the radio nor GeV band.  From the data we have, it appears that the broad-band spectra of very young pulsars have dominant hard X-ray components that become much less dominant relative to the GeV emission as pulsars age.  The transition between X-ray and GeV components lies in the MeV band.
\linebreak
\linebreak
Thus, even after the amazing advances from {\it Fermi}, we are left with a series of questions: How do RPP spectra evolve with age?  Where are the pairs produced in pulsar magnetospheres and how does their production evolve?  If pairs are produced near the polar caps, why don't we see their emission?  How large is the MeV pulsar population and why are they radio and GeV quiet?  What is the pulsar population near the Galactic center?  As we will detail below, telescopes sensitive in the 0.2--100\,MeV band can answer many of these questions, and the planned capabilities of proposed telescopes such as AMEGO (the All-sky Medium Energy Gamma-ray Observatory) \cite{Moiseev2017}, e-ASTROGAM \cite{DeAngelis2017}, AdEPT \cite{Hunter2014} and HARPO \cite{Gros2018} are well suited to the task.  If considered for the next decade, such telescopes will be primed to take advantage of the huge expansion in the population of radio RPPs expected from the Square Kilometre Array \cite{K2015}.

\section{Studying the Pulsar Life Cycle from MeV to GeV}

\begin{figure}[t]
\centering
\includegraphics[width=8cm]{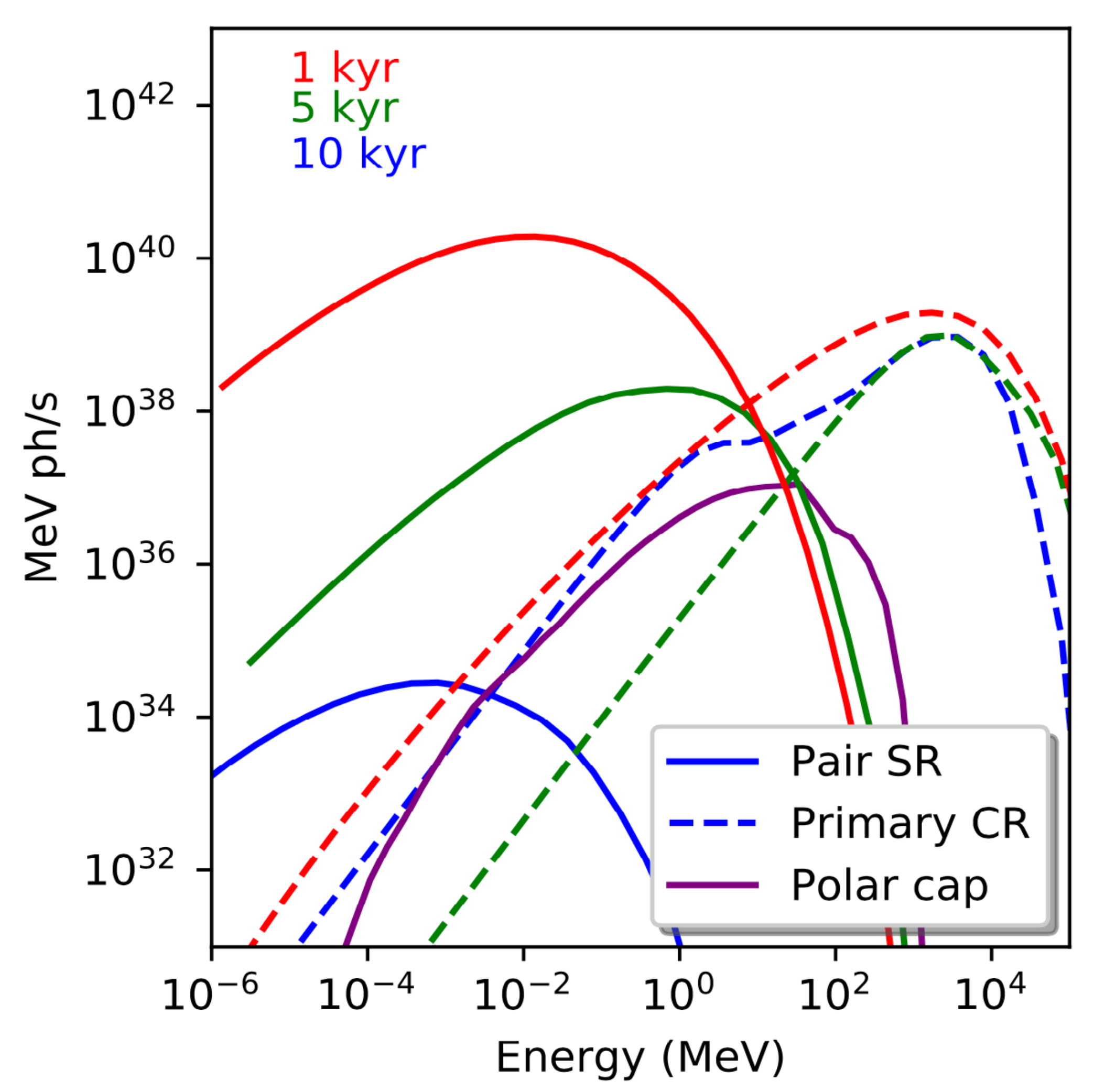}\includegraphics[width=8cm]{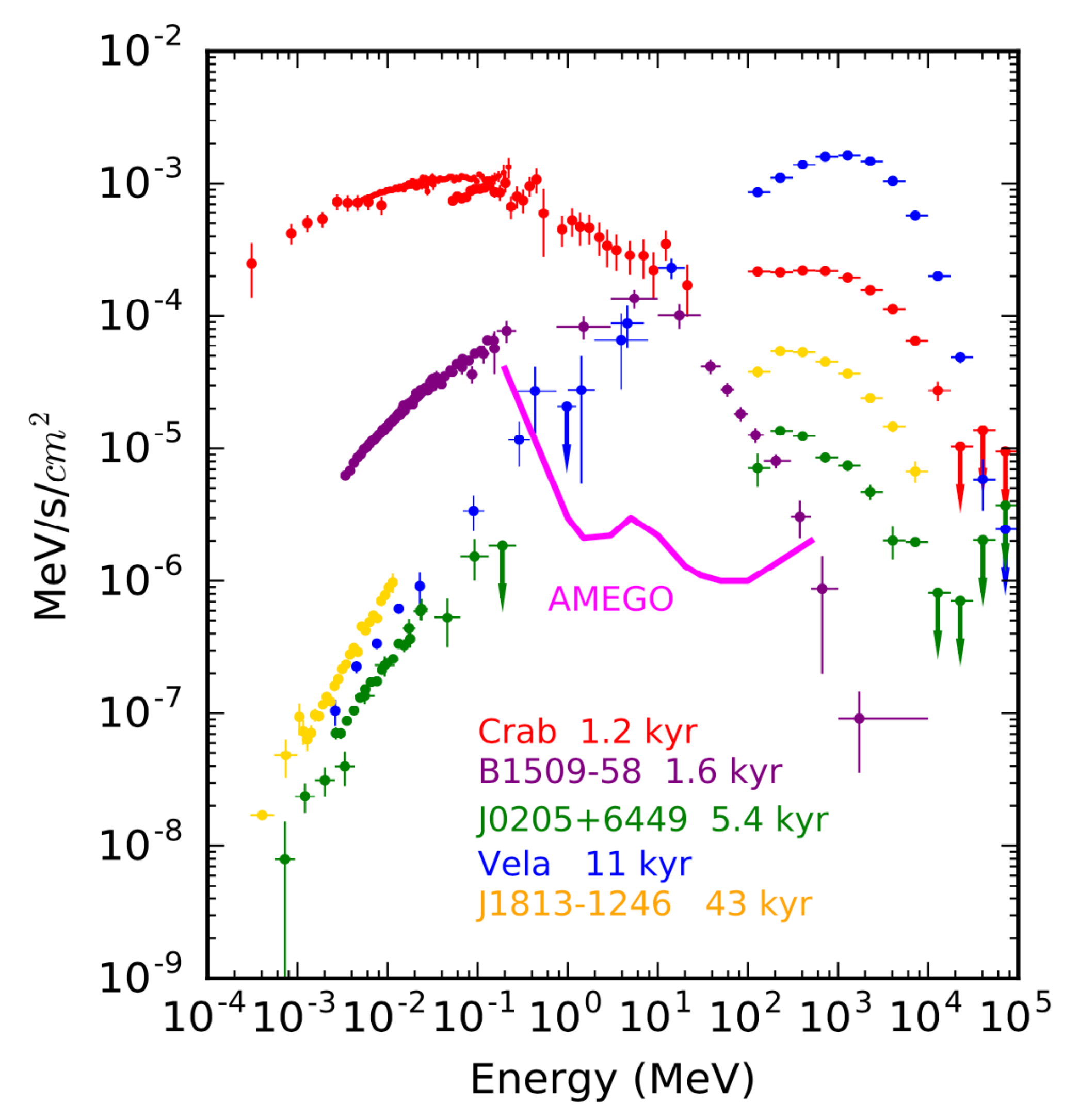}
\caption{\it {\bf Left panel:} Theoretical spectra for young pulsars \cite{HK2015,HK2018} showing strong evolution of the spectra as the pair multiplicity decreases with characteristic age. Additionally, direct $\gamma$-ray emission from the polar cap is expected to peak at 10 MeV \cite{TH2015}.   {\bf Right panel:} X-ray and $\gamma$-ray data from a selection of pulsars with broadband detections, along with the AMEGO sensitivity (data from \cite{KH2015,K2018}).}
\end{figure}   

Young RPPs (1-10 kyr) with large spin-down power produce electron-positron pair cascades that are thought to take place at the polar caps (PCs) \cite{HK2015}, the outer gaps \cite{ZC2000}, or current sheets \cite{PS2018}.  The pairs produce synchrotron radiation (SR) in the outer magnetosphere that peaks in the hard X-ray to soft $\gamma$-ray band.  A smaller number of particles are accelerated to high energy ($\sim 10$ TeV) in the current sheet and emit curvature radiation (CR) or SR at a few GeV, but the copious pairs effectively screen the accelerating electric fields in most of the rest of the magnetosphere and thereby limit the GeV emission.  As these pulsars age, the pair multiplicity, SR, and electric field screening drop, dramatically increasing the ratio of GeV-to-X-ray emission. This spectral life cycle, based on a number of models \cite{HK2015,Takata2010}, is clearly illustrated in Figure 1.  This picture can explain why the GeV luminosity varies so weakly with age and spin-down rate ($\sim \dot E^{1/2}$), but we do not understand why the pair emission is so sensitive to age.  Indeed, the understanding we have is largely theoretical, since there are only a few pulsars for which we have this full portrait, from MeV to GeV (Figure 1, right). With the pair SR component expected to decrease with age, better sensitivity in the MeV band is needed to explore how this emission evolves.  The sensitivity of an AMEGO-like telescope will allow exquisite measurements of the MeV spectrum and polarization of these known broadband emitters as well as discovery of new MeV pulsars (see below).
\linebreak
\linebreak
We compared two pulsar population synthesis models \cite{Gonthier2011,KMR2017} that satisfy constraints from pulsar surveys and found them largely consistent.  Adopting the population of \cite{Gonthier2011}, and scaling the $\gamma$-ray efficiency as $\eta = 0.1(10^{34}\rm erg\,s^{-1}/\dot E)^{0.5}$ above $\dot E = 10^{34}\,\rm erg\,s^{-1}$ and $\eta = 0.1$ for smaller $\dot E$, we estimate that an MeV telescope with flux sensitivity of $\sim 10^{-6}\,\rm MeV\,cm^{-2}\,s^{-1}$ could detect a few hundred pulsars, primarily through their GeV components (which are broad and detectable below a GeV; see Figure 2).  On the other hand, the SR component from pairs scales as $\sim 10^{-3}\,\dot E$ \cite{HM2011} and we expect to detect a fraction, $\sim 10$\% with pair components peaking near 0.1--1 MeV.   {\it Fermi} has detected a strong trend for young RPPs, namely a decrease in the spectral energy distribution (SED) peak of the GeV component with $\dot E$.  This trend may explain why there are no $\gamma$-ray pulsars detected with $\dot E \lsim 10^{33}\,\rm erg\,s^{-1}$.  
Theoretical studies of expected $\gamma$-ray spectra using global simulations \cite{Kala2018} have found that the spectra cutoffs, or SED peaks, move below 100 MeV for $\dot E \sim 10^{33-34}$\,erg\,s$^{-1}$ as the maximum particle acceleration energy drops.  Therefore, detecting the $\gamma$-ray emission from these low-$\dot E$ pulsars requires a telescope with good sensitivity below 100 MeV.   We can roughly estimate that a telescope with flux sensitivity of $\sim 10^{-6}\,\rm MeV\,cm^{-2}\,s^{-1}$ between 10 and 100 MeV could discover about 10 pulsars with $\dot E \lsim 10^{33}\,\rm erg\,s^{-1}$. 
\linebreak
\linebreak
The open question of whether the GeV emission is CR or SR can be definitively addressed by polarimetry measurements in the MeV-to-GeV component transition region of 1 - 1000 MeV.  Models have predicted \cite{TC2007,HK2017} that SR should have a low phase-averaged polarization degree around 10\% - 20\% while CR should have a much higher polarization degree around 40\% - 60\%. Time projection chamber telescopes like AdEPT \cite{Hunter2014} and HARPO \cite{Gros2018}, that are capable of measuring polarization with pair production, would be able to detect such a polarization-degree transition as well as an expected change in position angle if CR is the GeV emission mechanism.  
\linebreak
\linebreak
If, as expected, many of the electron-positron pairs are produced in cascades by QED magnetic pair creation near the PCs, the accelerated particles and secondary pairs should emit CR and SR $\gamma$-rays \cite{DH1982}.  The PC $\gamma$-ray emission was originally proposed to be the main observed GeV emission, but the spectra and light curves observed by {\it Fermi} have ruled this out in favor of outer magnetosphere emission.  However, the PC emission is still expected and can be unambiguously identified by its pulse phase near the radio peak and by phase-resolved polarimetry.  A recent estimate of its spectrum \cite{TH2015} for a Vela-like pulsar, shown in Figure 1, cuts off below the {\it Fermi} band around 10 MeV.  It is therefore expected that an MeV telescope will detect the PC emission for the first time, a milestone that will establish the PC as a source of pairs and enable detailed study of the high-field microphysics near neutron stars.
\linebreak
\begin{wrapfigure}{t}{0.50\textwidth}
\centering
\vskip -1.0cm
\includegraphics[width=8cm]{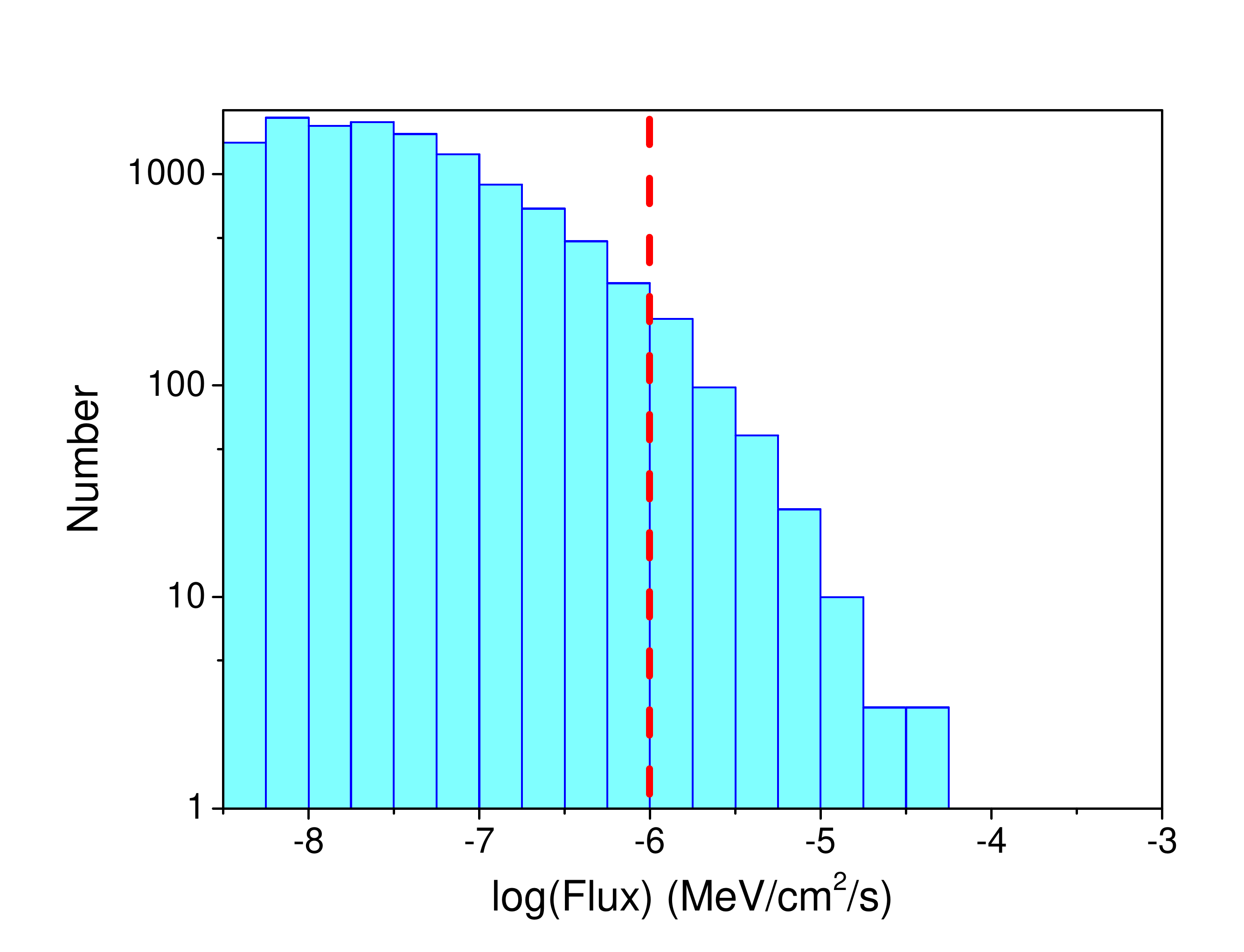}
\vskip -0.5cm
\caption{\it Estimated flux distribution of pulsars from a population synthesis model \cite{Gonthier2011}.  The red dashed line indicates the expected sensitivity limit of an AMEGO-like MeV telescope.}
\end{wrapfigure}   
Understanding the production of pairs by RPPs has broad implications.  The mechanism and evolution of pulsed radio emission from over 2500 RPPs is still not understood and constraining the multiplicity and spectrum of the PC pairs would provide a direct input to the the radio emission models.  The pair cascades also provide the particles that radiate broadband emission from radio to very-high-energy $\gamma$-rays in pulsar wind nebulae and the pair multiplicity is crucial to models of acceleration and evolution in these sources.  Pairs from pulsars have also been invoked to explain the observed local cosmic ray positron excess \cite{Hooper2009} and ultra-high-energy cosmic rays.

\section{The~Mystery~of~MeV~Pulsars}

A significant sub-population of young and energetic RPPs show high fluxes of pulsed non-thermal X-rays but no detected radio or GeV emission.  These so-called ``MeV pulsars" present a mystery since most pulsars with high spin down power have pulsed $\gamma$-ray and radio emission.  Is their anomaly due to geometry, that our line-of-sight misses both radio and GeV beams \cite{HK2017,Wang2014}, or to intrinsic weakness of acceleration and radio emission?  
Intriguingly, of the 18 known RPPs with X-ray emission above 20 keV \cite{KH2015}, 8 have narrow, double-peaked hard X-ray profiles and GeV emission while 10 (MeV pulsars) have broad, single-peaked hard X-ray profiles and no GeV emission.  Only one MeV pulsar has a detected spectral power peak, PSR B1509-58 at around 3 MeV, and one other, PSR J1846-0258 has a {\it Fermi}-LAT detection \cite{K2018} that implies a peak around 5 MeV (see Figure 3).  Many of the others have possible spectral peaks at least one order of magnitude above an AMEGO-like sensitivity.

\begin{wrapfigure}{r}{0.50\textwidth}
\centering
\vskip -0.2cm
\includegraphics[width=8cm]{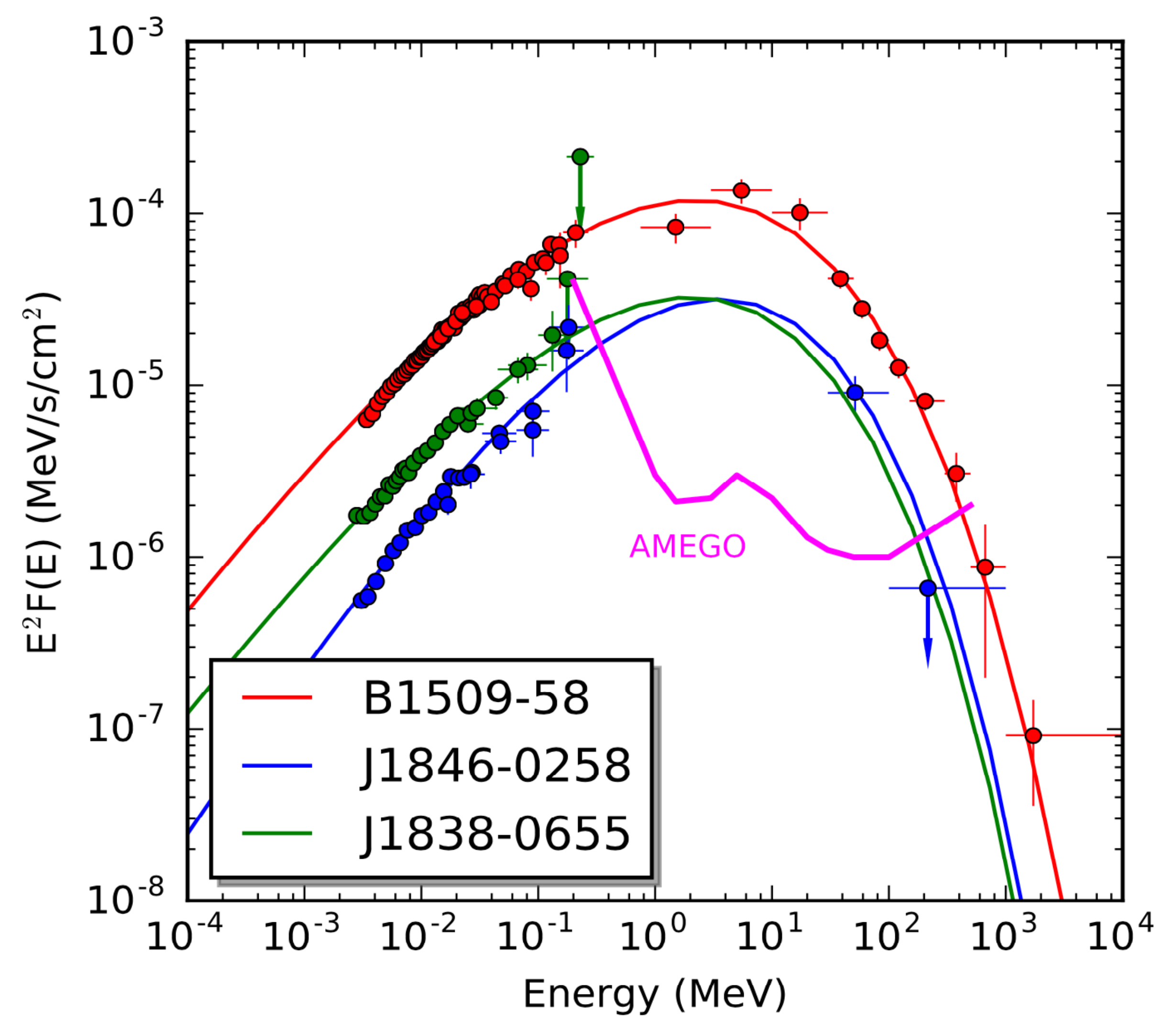}
\caption{\it A selection of MeV pulsars with broadband detections, along with the AMEGO sensitivity (data from \cite{KH2015,K2018}).}
\end{wrapfigure}

Measuring the properties of the MeV pulsar spectra can determine the location and constrain the spectrum and density of the pairs that are the prime source of plasma for the pulsar magnetosphere.  
The peak of the SED of SR from pairs in the outer magnetosphere would be $E_{\rm syn} \sim \gamma_+^2\,B_{\rm LC}$, where $\gamma_+$ is the maximum energy of the pairs, and $B_{\rm LC}= B_0 (R_0/R_{\rm LC})^3$ is the magnetic field at the light cylinder, $R_{\rm LC} = c/\Omega$, with $\Omega$ the pulsar rotation rate.  In models where polar cap pairs produce the SR, there is a strong dependence of $\gamma_+$, and therefore the SED peak, on surface magnetic field strength, $B_0$, $E_{\rm syn} \propto B_0 B_{LC}$ \cite{HK2017}. On the other hand, the SED peak of SR from outer gap and current sheet pairs has a stronger dependence on $B_{\rm LC}$, $E_{\rm syn} \propto B_{\rm LC}^{7/2}$ \cite{ZC2000}. For example, measuring the spectral power peaks of PSRs J1846-0258 and J1838-0655 will provide a strong discrimination between these models since both their values of $B_0 = 9 \times 10^{13}$ G and $3.8 \times 10^{12}$ G, and of $B_{\rm LC} = 2.5 \times 10^4$ G and $1.0 \times 10^5$ G are very different.
Discovering more MeV pulsars with a telescope like AMEGO, whose sensitivity matches their power peak, is very likely and could double the existing population \cite{DeAngelis2017}.  The shape of the light curves of a larger population would show whether source geometry is the cause of their GeV and radio silence.

\section{Galactic Center Pulsars}

An important legacy from \textit{Fermi} is the ``GeV excess'' \cite{Hooper2009,D2016,Ack2017}, a GeV-peaked, degree-scale signal at the Galactic center in excess of expectations from known diffuse and point sources.  Various explanations advanced include annihilating dark matter \cite{C2015} and cumulative emission from an unresolved population of millisecond \cite{BK2015,YZ2014} or young pulsars \cite{O2015}.  Finding these pulsars is challenging with radio searches due to interstellar scattering and dispersion.
\linebreak
\linebreak
The young pulsar hypothesis is particularly appealing: there is a deficit of millisecond pulsar progenitors \cite{H2017}, and the detection of a few radio pulsars suggests a population of up to $\sim 1000$ RPPs with $\dot E \gsim 10^{35}\,\rm erg\,s^{-1}$ near the Galactic center \cite{Deneva2009,Wharton2012}.  In addition, the high star formation rate there (e.g. Arches cluster) suggests that a few very energetic newborn neutron stars lie within 100 pc of Sgr A*.  As we show below, an MeV telescope could definitively detect both these newborn pulsars and a substantial portion of this population.\\

\begin{wrapfigure}{r}{0.50\textwidth}
\centering
\vskip -0.2cm
\includegraphics[width=8cm]{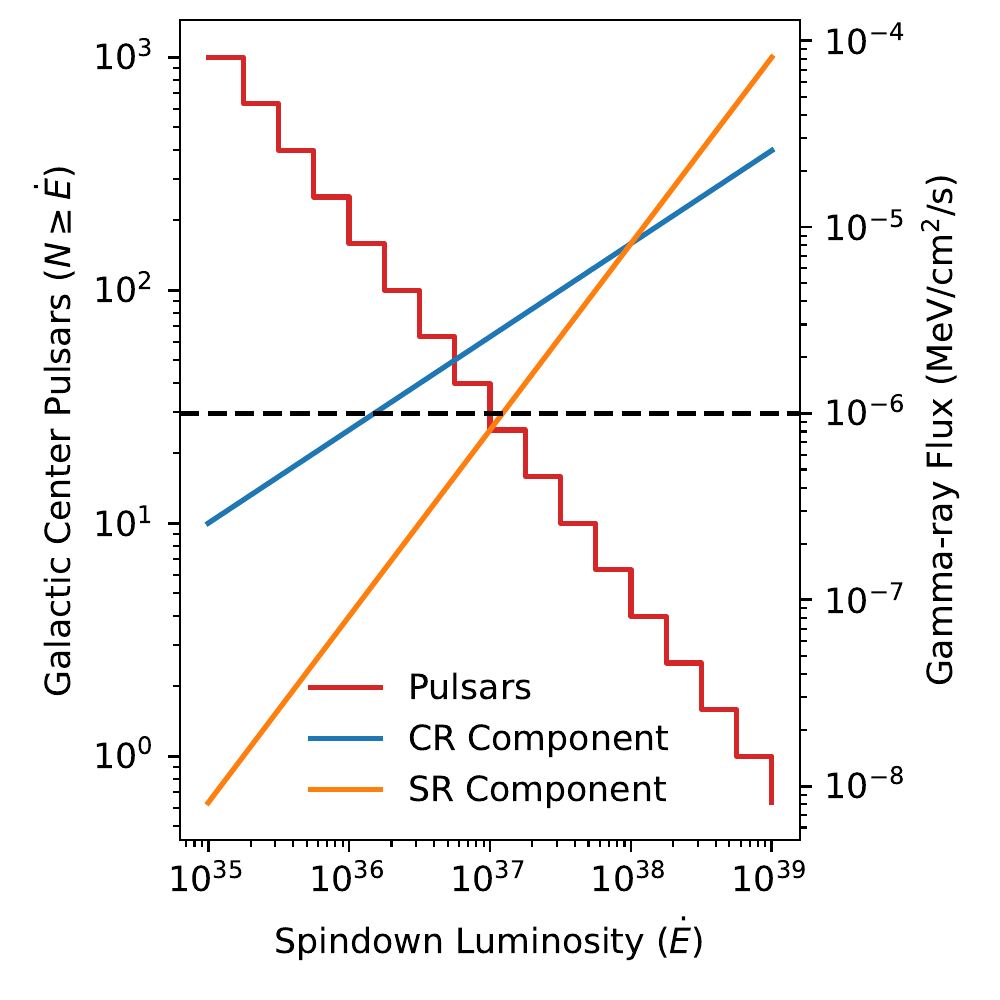}
\caption{\it \textbf{Left:} The Galactic center population using an estimated shape of the population from synthesis models \cite{KMR2017} and normalization from \cite{Deneva2009,Wharton2012}. \textbf{Right:} The $\gamma$-ray flux for two radiation components as described in \S4.  The notional AMEGO sensitivity is shown by the dashed black line.}
\end{wrapfigure}

Using the pulsar population synthesis models introduced in \S2, we determined that the cumulative distribution of high-$\dot E$ pulsars ($\dot E>10^{35}$\,erg\,s$^{-1}$) is well described by a power law $N(\geq\dot E)\sim\dot E^{-\frac{4}{5}}$, and we show this population and the following flux estimates in Figure 4.  Because the SR luminosity from pairs scales as $10^{-3}\dot E$ (see \S2), the most energetic pulsars are preferentially detected through this channel.  With a sensitivity of $10^{-6}$\,MeV\,cm$^{-2}$\,s$^{-1}$, the $\dot{E}$ threshold for SR at 8 kpc is $10^{37}$\,erg\,s$^{-1}$.  For a population of 1000 pulsars, about 3\% satisfy this threshold, and an MeV telescope like AMEGO could discover 30 energetic pulsars near the Galactic center, firmly establishing the presence of young neutron stars.  Moreover, because newborn pulsars generally satisfy $\dot E>10^{37}$\,erg\,s${-1}$, AMEGO would strongly detect any young (few kyr) pulsars produced by the intense star formation near Sgr A*.  Although the primary (CR) component is in principle detectable for pulsars with $\dot E>10^{36}$\,erg\,s$^{-1}$, \textit{Fermi} could not necessarily have detected members of this population, as its angular resolution is only $\sim$1$^{\circ}$ at 1\,GeV.  An MeV telescope relying on an all-silicon tracker will have a substantially better angular resolution \cite{Moiseev2017} and thus be better able to detect sources through either channel.  Indeed, this may be critical, because ``blindly'' detecting pulsations requires accumulating many source photons as quickly as possible in order to overcome timing noise.
\linebreak
\linebreak
Finally, an MeV telescope could also detect a population of millisecond pulsars (MSPs).  Although less energetic than unrecycled pulsars, MSPs may be particularly efficient emitters of MeV $\gamma$-rays \cite{HM2011,HK2015}.  Indeed, recent observations of hard X-ray emission from three energetic MSPs \cite{GB2017} suggest the bulk of their energy is emitted in the MeV band.  If their efficiencies surpass 1\%, then MSPs like B1821$-$24A and B1937$+$21, with $\dot{E}>10^{36}$\,erg\,s$^{-1}$, could be detected at the Galactic center and signal the presence of a larger population.  Detection of either population of pulsars would be a major step forward in understanding both the ``GeV excess'' and the Galactic center more generally.




\bibliographystyle{plain}

\end{document}